\documentstyle [12pt] {article}
\begin{document}
\setcounter{page}{1}
\def\theequation{\arabic{section}.\arabic{equation}}
\def\theequation{\thesection.\arabic{equation}}
\setcounter{section}{0}

\vspace*{1.0cm}
\begin{center}
\section*{\bf On the S--wave $\pi$D--scattering length\\ in the
relativistic field theory model of the
deuteron}
\smallskip

{\bf A. N. Ivanov\footnote{\normalsize E--mail:
ivanov@kph.tuwien.ac.at,Tel.:+43--1--58801--5598,
Fax:+43--1--5864203}$^{,2}$, N.I.
Troitskaya\footnote{\normalsize Permanent Address:
State Technical University, Department of Theoretical
Physics, 195251 St. Petersburg, Russian Federation.},
M.Faber\footnote{\normalsize E--mail: faber@kph.tuwien.ac.at,
Tel.:+43--1--58801--5598, Fax:+43--1--5864203}, H.
Oberhummer\footnote{\normalsize E--mail: ohu@kph.tuwien.ac.at,
Tel.:+43--1--58801--5574, Fax:+43--1--5864203}}
\vskip1.0truecm

{\it Institut f\"ur Kernphysik, Technische Universit\"at Wien,}\\
\baselineskip=14pt
{\it Wiedner Hauptstr. 8-10, A-1040 Vienna, Austria}\\
\end{center}

\begin{center}
\begin{abstract}
The S--wave scattering length of the strong pion--deuteron ($\pi$D)
scattering is calculated in the relativistic field theory model of the
deuteron suggested in [1,2].The theoretical result agrees well with the
experimental data. The important role of the $\Delta$--resonance
contribution to the elastic $\pi$D--scattering is confirmed.
\end{abstract}
\end{center}

\begin{center}
PACS: 11.10. Ef, 12.39. Fe, 13.75. Cs, 13.75 Gx \\
\noindent Keywords: deuteron, pion, chiral symmetry, pion--baryon
interaction, $\Delta$--resonance, phenomenological model
\end{center}
\vskip1.0truecm

\newpage

\section{Introduction}
\setcounter{equation}{0}

In our previous publications [1,2] we have suggested a relativistic field
theory model of the deuteron. The basis of the model is the one--nucleon
loop origin of a physical deuteron produced by
low--energy fluctuations of the proton and neutron.
Therefore, the model can be applied to the description of only low--energy
interactions of the deuteron. All interactions of a physical deuteron with
other
particles should be determined only by one--nucleon loops [1,2]. This
recipe
is very similar to the approximation accepted within the
Nambu--Jona--Lasinio model, where all interactions of hadrons, collective
quark--antiquark excitations, are obtained through one--constituent
quark--loop exchanges [3-7]. Also one has to understand that most likely
the deuteron
cannot be inserted in an intermediate state of any process of low--energy
interactions.
This is connected with a very sensitive structure of the deuteron as an
"extended" bound state with a small binding energy. The representation of
such a state in terms of any local quantum field is rather limited. The
latter entails an undetermined character of the description of the deuteron
in intermediate states in terms
of Green functions of these local fields.

In this paper we apply such a model to the computation of the S-wave
elastic $\pi$D--scattering.

The effective Lagrangian of the physical deuteron field $D_{\mu}(x)$
interacting strongly with the proton $p(x)$, neutron $n(x)$ and pion
$\vec{\pi}(x)$
fields reads [1,2]
\begin{eqnarray}\label{label1}
{\cal L}_{\rm st}(x) & = & -\frac{1}{2} D^{\dagger}_{\mu\nu}(x)
D^{\mu\nu}(x) + M^2_{\rm D} D^{\dagger}_\mu(x) D^{\mu}(x) +
\nonumber\\
&& + g_{\rm V}\bar{N}(x) \gamma^\mu \tau^2 N^c(x) D_\mu (x) + g_{\rm
V}\bar{N^c}(x)
\tau^2 \gamma^\mu N(x) D^{\dagger}_\mu (x) + \\
&& + {\cal L}_{\rm \pi NN}(x)\,,\nonumber
\end{eqnarray}
where $\vec{\tau}=(\tau^1, \tau^2, \tau^3)$ are Pauli matrices of the
isotopical spin,
and $N(x)$ is the nucleon field, the isotopical doublet with components
$N(x) = (p(x),
n(x))$. Under isotopical rotations the field $\tau_2N^c(x)$ transforms like
$N(x)$.
Then $D_{\mu\nu}(x)=\partial_{\mu} D_{\nu}(x)-\partial_{\nu} D_{\mu}(x)$,
$M_{\rm D}=2M_{\rm N} - \varepsilon_{\rm D}$ is the mass of the physical
deuteron,
where $M_{\rm N} = 938\,{\rm MeV}$ is the mass of the proton and neutron
and
$\varepsilon_{\rm D} = 2.225\,{\rm MeV}$ is the binding energy of the
deuteron [8],
$g_{\rm V}$ is the phenomenological coupling constant of the model [2],
that can be
expressed in terms of the electric quadrupole moment of the deuteron
$Q_{\rm D}$ [2], i.e., $g^2_{\rm V}/\pi^2 = Q_{\rm D} M^2_{\rm D}/2$.
As $Q_{\rm D}=0.286\,{\rm fm}^2$ [8] we get $g_{\rm V}=11.3$ [2].
Then $\psi^c(x) = C \bar{\psi}^T(x)$
and $\bar{\psi^c}(x) = \psi^T(x) C$, where the indices $c$ and $T$ imply a
charge
conjugation and a transposition, and $C = i\gamma^2\gamma^0$ is the matrix
of a charge conjugation.

The effective Lagrangian ${\cal L}_{\rm \pi NN}(x)$ describes strong
low--energy interactions of neucleons and pions. Since it is well--known
that chiral $SU(2)\times SU(2)$ symmetry plays an important role for strong
low--energy interactions of pions and nucleons, we should use the
pion--nucleon interactions invariant under chiral $SU(2)\times SU(2)$
transformations.

There are two possibilities for the realization of chiral $SU(2)\times
SU(2)$ symmetry, i.e. using either linear or nonlinear representation. In
a linear
realization of chiral $SU(2)\times SU(2)$ symmetry one needs to introduce a
scalar
isoscalar field $\sigma(x)$ with the mass of the order
$M_{\sigma}\sim 700\,{\rm MeV}$
being a partner of a pion field under chiral rotations [4-7].
Unfortunately, up
to now the scalar isoscalar meson with the mass $M_{\sigma}\sim 700\,{\rm
MeV}$ was
not observed experimentally. In turn, within a nonlinear realization of
chiral
$SU(2)\times SU(2)$ symmetry, a pion field transforms nonlinearly under
chiral
rotations and the scalar isoscalar field $\sigma(x)$ does not appear. To
avoid
the problem of dealing with unobserved state we suggest to use a nonlinear
realization of
chiral $SU(2)\times SU(2)$ symmetry. In this case, following Weinberg
[9], the Lagrangian ${\cal L}_{\rm \pi NN}(x)$ should read
\begin{eqnarray}\label{label2}
&&{\cal L}_{\rm \pi NN}(x) =\bar{N}(x) (i\gamma^\mu \partial_\mu - M_{\rm
N}) N(x) -\Bigg[1+\frac{\vec{\pi}^2(x)}{4
F^2_0}\Bigg]^{-1}\times\nonumber\\
&&\Bigg[\frac{g_{\rm \pi NN}}
{2 M_{\rm N}}\bar{N}(x) \gamma^{\mu} \gamma^5 \vec{\tau} N(x) \cdot
\partial_{\mu}\vec{\pi}(x) + \frac{1}{4 F^2_0}\,\bar{N}(x)\gamma^{\mu}
\vec{\tau}N(x) \cdot (\vec{\pi}(x)
\times \partial_{\mu}\vec{\pi}(x))\Bigg]\nonumber\\
&&+ \frac{1}{2}\partial_{\mu}\vec{\pi}(x)\cdot \partial^{\mu}\vec{\pi}(x)\,
\Bigg[1+\frac{\vec{\pi}^2(x)}{4 F^2_0}\Bigg]^{-1} - \frac{1}{2}M^2_{\pi}
\vec{\pi}^2(x)\,
\Bigg[1+\frac{\vec{\pi}^2(x)}{4 F^2_0}\Bigg]^{-1}\,=\\
&&= \bar{N}(x) (i\gamma^\mu \partial_\mu - M_{\rm N}) N(x) +
\frac{1}{2}\partial_{\mu}\vec{\pi}(x)\cdot \partial^{\mu}\vec{\pi}(x)\,
- \frac{1}{2}M^2_{\pi} \vec{\pi}^2(x)\,-\nonumber\\
&&- \frac{g_{\rm \pi NN}}
{2 M_{\rm N}}\bar{N}(x) \gamma^{\mu} \gamma^5 \vec{\tau} N(x) \cdot
\partial_{\mu}\vec{\pi}(x) - \frac{1}{4 F^2_0}\,\bar{N}(x)\gamma^{\mu}
\vec{\tau}N(x) \cdot (\vec{\pi}(x)\times
\partial_{\mu}\vec{\pi}(x))\nonumber\\
&& + \ldots ,\nonumber
\end{eqnarray}
where $M_{\pi}=M_{\pi^0}=134.976\,{\rm MeV}$ is the mass of the pion field
$\vec{\pi}(x)$, $F_0\simeq 92\,{\rm MeV}$ is
the PCAC constant of $\pi$--mesons calculated in the chiral limit [7],
and $g_{\rm \pi NN}$ is the ${\rm \pi NN}$--coupling constant.
Below we set $g_{\rm \pi NN}=13.4\pm 0.2$ [2,10].

However,
it is well--known [11] that the ${\rm \pi NN}$--interaction is not
sufficient to explain strong low--energy interactions of the deuteron, and
the interaction of pions and nucleons with the $\Delta$--resonance, i.e.
the ${\rm \pi N \Delta}$--interaction, plays an important role. Following
[12], we take the ${\rm \pi N \Delta}$--interaction in the form
\begin{eqnarray}\label{label3}
{\cal L}_{\rm \pi N \Delta}(x) = \frac{g_{\rm \pi N \Delta}}{2
M_N}\,\bar{\Delta}^a_{\mu}(x)\,N(x)\,\partial^{\mu}\pi^a(x) + {\rm h.c.} ,
\end{eqnarray}
where $\Delta^a_{\mu}(x)$ is the $\Delta$--resonance field, the index $a$
runs over $a=1,2,3$ and

\parbox{11cm}{\begin{eqnarray*}
\begin{array}{llcl}
&&\Delta^1=\,\frac{1}{\sqrt{2}}\Biggr(\begin{array}{c}
\Delta^{++}-\Delta^0/\sqrt{3} \\ \Delta^+/\sqrt{3} - \Delta^-
\end{array}\Biggl)\,,\,
\Delta^2=\,\frac{i}{\sqrt{2}}\Biggr(\begin{array}{c}
\Delta^{++}+\Delta^0/\sqrt{3} \\ \Delta^+/\sqrt{3} + \Delta^-
\end{array}\Biggl)\,,\\
&&\Delta^3=\,-\sqrt{\frac{2}{3}}\Biggr(\begin{array}{c} \Delta^+ \\
\Delta^0 \end{array}\Biggl)\,.
\end{array}
\end{eqnarray*}} \hfill
\parbox{1cm}{\begin{eqnarray}\label{label4}
\end{eqnarray}}

\noindent Using the ${\rm \pi N \Delta}$--interaction (\ref{label3}) we
derive the effective Lagrangian describing the $\pi$N $\to \Delta \to$
$\pi$N transition
\begin{eqnarray}\label{label5}
&&\int d^4x {\cal L}^{\pi{\rm N}\to \Delta \to \pi{\rm N}}_{\rm
eff}(x)=\frac{ig^2_{\rm \pi N \Delta}}{4 M^2_N}
\int d^4x_1 d^4x_2\,\times\\
&&\times\,\Big[\bar{N}(x_1)<0|{\rm T}
\Big(\Delta^a_{\mu}(x_1)\bar{\Delta}^b_{\nu}(x_2)\Big)|0>N(x_2)\Big]\,
\partial^{\mu}\pi^a(x_1)\partial^{\nu}\pi^b(x_2),\nonumber
\end{eqnarray}
where
\begin{eqnarray}\label{label6}
<0|{\rm
T}\Big(\Delta^a_{\mu}(x_1)\bar{\Delta}^b_{\nu}(x_2)\Big)|0>=-i\Bigg(\frac{2}
{3}\delta^{ab}-\frac{1}{3}i\varepsilon^{abc}\tau^c\Bigg)S_{\mu\nu}(x_1-x_2).
\end{eqnarray}
We define the momentum representation of $S_{\mu\nu}(x)$
in accordance with
[13]
\begin{eqnarray}\label{label7}
S_{\mu\nu}(p) = \frac{1}{M_{\Delta} - \hat{p}}\,\Bigg(- g_{\mu\nu} +
\frac{1}{3} \gamma_{\mu}\gamma_{\nu} +
\frac{1}{3}\,\frac{\gamma_{\mu}p_{\nu}-\gamma_{\nu}p_{\mu}}{M_{\Delta}} +
\frac{2}{3}\,\frac{p_{\mu}p_{\nu}}{M^2_{\Delta}}\Bigg),
\end{eqnarray}
where $M_{\Delta}=1232\;{\rm MeV}$ is the $\Delta$--resonance mass.

Substituting (\ref{label6}) in (\ref{label5}) we get
\begin{eqnarray}\label{label8}
&&\int d^4x {\cal L}^{\pi{\rm N}\to \Delta \to \pi{\rm N}}_{\rm eff}(x)=\\
&&=\frac{g^2_{\rm \pi N \Delta}}{6 M^2_N}
\int d^4x_1
d^4x_2\times\Big[\bar{N}(x_1)S_{\mu\nu}(x_1-x_2)N(x_2)\Big]\,\partial^{\mu}\
\vec{\pi}(x_1)\cdot \partial^{\nu}\vec{\pi}(x_2) -\nonumber\\
&&-\frac{g^2_{\rm \pi N \Delta}}{12 M^2_N}\int d^4x_1 d^4x_2
\Big[\bar{N}(x_1)S_{\mu\nu}(x_1-x_2)i\vec{\tau}\,N(x_2)\Big]\cdot(\partial^{
\mu}\vec{\pi}(x_1)\times \partial^{\nu}\vec{\pi}(x_2)).\nonumber
\end{eqnarray}
The coupling constant $g_{\rm \pi N \Delta}$ is connected with $g_{\rm
\pi NN}$ via the relation $g_{\rm \pi N \Delta} = 1.90 \, g_{\rm \pi
NN}$ [12,14].

We should emphasize that we do not take into account the ${\rm \pi
\Delta \Delta}$--interaction. It is because in our model defined in
one--baryon loop approximation the contribution of the ${\rm \pi
\Delta \Delta}$--interaction is the matter of two--baryon loop
approximation.

\section{Elastic $\pi$D--scattering}
\setcounter{equation}{0}

The effective Lagrangian ${\cal L}^{\rm \pi D \to \pi D}_{\rm eff}(x)$,
describing elastic $\pi$D--scattering at low energies, contains  two
contributions
\begin{eqnarray}\label{label9}
{\cal L}^{\rm \pi D \to \pi D}_{\rm eff}(x)={\cal L}^{\rm \pi D \to NN \to
\pi D}_{\rm eff}(x) + {\cal L}^{\rm \pi D \to N\Delta \to \pi D}_{\rm
eff}(x),
\end{eqnarray}
where ${\cal L}^{\rm \pi D \to NN \to \pi D}_{\rm eff}(x)$ and ${\cal
L}^{\rm \pi D \to N\Delta \to \pi D}_{\rm eff}(x)$ are represented by the
simple box--diagrams and defined
\begin{eqnarray}\label{label10}
&&\int\,d^4x\,{\cal L}^{\rm \pi D \to NN \to \pi D}_{\rm
eff}(x)\,=\int\,d^4x\,\int\,
\frac{d^4x_1\,d^4k_1}{(2\pi)^4}\,\frac{d^4x_2\,d^4k_2}{(2\pi)^4}\,\frac{d^4x
_3\,
d^4k_3}{(2\pi)^4}\times\nonumber\\
&& \times
D_{\mu}(x)\,D^{\dagger}_{\nu}(x_1)\,\partial_{\alpha}\vec{\pi}(x_2)
\cdot \partial_{\beta}\vec{\pi}(x_3) \times \\
&&\times e^{-ik_1\cdot x_1} e^{-ik_2\cdot x_2} e^{-ik_3\cdot
x_3}e^{i(k_1+k_2+k_3)
\cdot x}\,\frac{g^2_{\rm V}}{8\pi^2}\frac{g^2_{\rm \pi NN}}{4M^2_{\rm N}}\,
{\cal J}^{\mu\nu\alpha\beta}(k_1, k_2, k_3; Q)_{\rm NN}\,,\nonumber
\end{eqnarray}
\begin{eqnarray}\label{label11}
&&\int\,d^4x\,\tilde{{\cal L}}^{\rm \pi D \to N\Delta \to \pi D}_{\rm
eff}(x)\,=\int\,d^4x\,\int\,
\frac{d^4x_1\,d^4k_1}{(2\pi)^4}\,\frac{d^4x_2\,d^4k_2}{(2\pi)^4}\,\frac{d^4x
_3\,
d^4k_3}{(2\pi)^4}\times\nonumber\\
&& \times
D_{\mu}(x)\,D^{\dagger}_{\nu}(x_1)\,\partial_{\alpha}\vec{\pi}(x_2)
\cdot \partial_{\beta}\vec{\pi}(x_3) \times \\
&&\times e^{-ik_1\cdot x_1} e^{-ik_2\cdot x_2} e^{-ik_3\cdot
x_3}e^{i(k_1+k_2+k_3)
\cdot x}\,\frac{g^2_{\rm V}}{8\pi^2}\frac{g^2_{\rm \pi N \Delta}}{6
M^2_{\rm N}}
{\cal J}^{\mu\nu\alpha\beta}(k_1, k_2, k_3; Q)_{\rm N\Delta}.\nonumber
\end{eqnarray}
The structure functions ${\cal J}^{\mu\nu\alpha\beta}(k_1, k_2, k_3;
Q)_{\rm NN}$ and ${\cal J}^{\mu\nu\alpha\beta}(k_1, k_2, k_3; Q)_{\rm
N\Delta}$ are given by

\parbox{11cm}{\begin{eqnarray*}
&&{\cal J}^{\mu\nu\alpha\beta}(k_1, k_2, k_3; Q)_{\rm
NN}=\int\,\frac{d^4k}{\pi^{\,2}\,i}
{\rm tr}\,\{S_{\rm F}(k+Q) \gamma^{\mu}S_{\rm F}(k+Q+k_1) \gamma^{\nu}\\
&&S_{\rm F}(k+Q+k_1+k_2)\gamma^{\alpha}\gamma^5
S_{\rm F}(k+Q+k_1+k_2+k_3)\gamma^{\beta}
\gamma^5\}\,,\\
&&{\cal J}^{\mu\nu\alpha\beta}(k_1, k_2, k_3; Q)_{\rm N\Delta}=
\int\,\frac{d^4k}{\pi^{\,2}\,i}
{\rm tr}\,\{S_{\rm F}(k+Q) \gamma^{\mu} S_{\rm F}(k+Q+k_1)\gamma^{\nu}\\
&& S_{\rm F}(k+Q+k_1+k_2) S^{\alpha\beta}(k+Q+k_1+k_2+k_3)\}\,,
\end{eqnarray*}} \hfill
\parbox{1cm}{\begin{eqnarray}\label{label12}
\end{eqnarray}}

\noindent where $S_{\rm F}(p)$ is a Green function of a free nucleon in the
momentum representation
\begin{eqnarray}\label{label13}
S_{\rm F}(p) =\frac{1}{M_{\rm N} - \hat{p}}\;.
\end{eqnarray}
Then $Q = a\,k_1 + b\,k_2 + c\,k_3$ is an arbitrary shift of a virtual
momentum.
Fortunately, the result of the computation of the integrals (\ref{label12})
taken in leading long--wavelength approximation [1,2] does not depend on
the
shift of a virtual momentum and can be computed unambiguously. Holding the
terms giving the main contribution in long--wavelength expansion [1,2] we
obtain

\parbox{11cm}{\begin{eqnarray*}
{\cal J}^{\mu\nu\alpha\beta}(k_1, k_2, k_3; Q)_{\rm
NN}&=&-\frac{16}{9}\,g^{\mu\nu}\,g^{\alpha\beta}+\ldots,\\
{\cal J}^{\mu\nu\alpha\beta}(k_1, k_2, k_3; Q)_{\rm
N\Delta}&=&-\frac{20}{9}\,g^{\mu\nu}\,g^{\alpha\beta}+\ldots\,.
\end{eqnarray*}} \hfill
\parbox{1cm}{\begin{eqnarray}\label{label14}
\end{eqnarray}}

\noindent We have neglected the contribution of the mass difference
$M_{\Delta} - M_{\rm N}$ that is small compared with the remained part.
Also we have dropped out the divergent contributions that can be expressed
in terms of the integrals [1,2]

\parbox{11cm}{\begin{eqnarray*}
J_{\,1}(M_{\rm N}) &=&\int\frac{d^4k}{\pi^2i}\frac{1}{M_{\rm N}^2 - k^2}=
4\int^{\Lambda_{\rm D}}_{0}\frac{d|\vec{k}|{\vec{k}}^{\,2}}{(M^2_{\rm
N}\,+\,{\vec{k}}^{\,2})^{\,1/2}}\,,\\
J_{\,2}(M_{\rm N}) &=&\int\frac{d^4k}{\pi^2i}\frac{1}{(M_{\rm N}^2 -
k^2)^{\,2}}= 2\int^{\Lambda_{\rm
D}}_{0}\frac{d|\vec{k}|{\vec{k}}^{\,2}}{(M^2_{\rm
N}\,+\,{\vec{k}}^{\,2})^{\,3/2}}\,.
\end{eqnarray*}} \hfill
\parbox{1cm}{\begin{eqnarray}\label{label15}
\end{eqnarray}}

\noindent The ultraviolet cut--off has the meaning of the upper limit
$\Lambda_{\rm D} =
64.843\,{\rm MeV}$ restricting the 3--momenta of fluctuations of virtual
nucleons
taking part in the formation of the physical deuteron field [1,2].
In terms of the cut--off $\Lambda_{\rm D}=64.843\;{\rm MeV}$ we define an
effective radius of the deuteron $r_{\rm D}= 1/\Lambda_{\rm D}= 3.043\;{\rm
fm}$ [2]. This value agrees well with the average value of the deuteron
radius $\bar{r} = \int d^3x\,r\,\vert \psi_{\rm D}(\vec{r}\,)\vert^2 =
3.140\;{\rm fm}$ [15], where $\psi_{\rm D}(\vec{r}\,)$ is the wave--function of
the deuteron at the ground state. The effective radius  $r_{\rm D}=
1/\Lambda_{\rm D}= 3.043\;{\rm fm}$ is compared qualitatively with the deuteron
radius $r_{\rm D}= (\varepsilon_{\rm D}M_{\rm N})^{-1/2} = 4.319\;{\rm fm}$
[8], defined in terms of the binding energy of the deuteron $\varepsilon_{\rm
D} = 2.225\;{\rm MeV}$ [8].
Due to
inequality
$M_{\rm N} \gg \Lambda_{\rm D}$ the term depending on the cut--off
$\Lambda_{\rm D}$
is small compared with convergent contributions.

Applying structure functions Eq.(\ref{label14}) we compute the effective
Lagrangian describing the elastic low--energy $\pi$D--scattering
\begin{eqnarray}\label{label16}
{\cal L}^{\rm \pi D \to \pi D}_{\rm eff}(x)&=&-\frac{1}{9}\,Q_{\rm
D}\,\Bigg(g^2_{\rm \pi NN}+\frac{5}{6}\,g^2_{\rm \pi N\Delta}\Bigg)
\,D^{\dagger}_{\mu}(x)\,D^{\mu}(x)\,\partial_{\nu}\vec{\pi}(x) \cdot
\partial^{\nu}\vec{\pi}(x) .
\end{eqnarray}
The amplitude of the elastic low--energy $\pi$D--scattering defined by the
effective Lagrangian (\ref{label14}) reads
\begin{eqnarray}\label{label17}
{\cal A}(s, t, u)\,=\,\frac{1}{9}\,Q_{\rm D}\,\Bigg(g^2_{\rm \pi
NN}+\frac{5}{6}\,g^2_{\rm \pi N\Delta}\Bigg)\,(t - 2 M^2_{\pi})\,.
\end{eqnarray}
In turn the S--wave length of the elastic $\pi$D--scattering is given by

\parbox{11cm}{\begin{eqnarray*}
a_{\rm \pi D}&=& \frac{1}{8\pi}\frac{1}{M_{\rm D} + M_{\pi}}{\cal A}(s, t,
u)\Bigg|_{\rm threshold}\,=\\
&=&-\frac{1}{36\pi}\,\Bigg(g^2_{\rm \pi NN}+\frac{5}{6}\,g^2_{\rm \pi
N\Delta}\Bigg)\frac{Q_{\rm D}M^2_{\pi}}{M_{\rm D}
+ M_{\pi}}\,=\,-\,0.057\,M^{-1}_{\pi}\,.
\end{eqnarray*}} \hfill
\parbox{1cm}{\begin{eqnarray}\label{label18}
\end{eqnarray}}

\noindent The S--wave scattering length Eq.(\ref{label18}) is of order
$O(M^2_{\pi})$. This agrees with the results obtained by Robilotta [16] and
Weinberg [17]. The theoretical magnitude of $a_{\rm \pi D}$ given by
Eq.(\ref{label18}) is reasonably well compared with experimental data [18]
\begin{eqnarray}\label{label19}
{(a_{\rm \pi D})}_{\exp}=-0.052^{+0.022}_{-0.017}\,M^{-1}_{\pi}.
\end{eqnarray}
One can see that the ${\rm \pi N\Delta}$--interaction gives the main
contribution to the low--energy S--wave elastic $\pi$D--scattering. This
confirms the important role of the $\Delta$--resonance for strong
low--energy $\pi$D--dynamics.

\section{Conclusion}

We have shown that the relativistic field theory model of the deuteron
supplemented by the effective pion--nucleon interaction derived within
nonlinear
realization of chiral $SU(2)\times SU(2)$ symmetry  and contribution of the
$\Delta$--resonance describes well the dynamics of
low--energy elastic $\pi$D--scattering at energies very close to the
$\pi$D--threshold.

The magnitude of the S--wave scattering length is found to be
well compared  with
experimental data due to the inclusion of the ${\rm \pi N
\Delta}$--interaction. This confirms a well--known assertion concerning an
important role of the $\Delta$--resonance in the elastic
$\pi$D--scattering.

Thus we can conclude that the relativistic field theory model of the
deuteron really can be applied to the description of strong low--energy
$\pi$D--dynamics at energies very close to the $\pi$D--threshold.

We acknowledge fruitful discussions with Prof.~G. E.~Rutkovsky and Dr. H.
Leeb. We thank Dr. L. G. Dakhno for reading manuscript and corrections.
This work was
partially supported by the Fonds zur F\"orderung wissenschaftlichen
Forschung in
\"Osterreich (project P10361--PHY).

\end{document}